\documentclass[9pt,twocolumn,twoside]{opticajnl}
\journal{opticajournal} 
\usepackage{hyperref}

\setboolean{shortarticle}{false}

\usepackage{lineno}

\title{Near-optimal decomposition of unitary matrices using \\ phase masks and the discrete Fourier transform}

\author[1,*]{Vincent Girouard}
\author[1,$\dagger$]{Nicol\'as Quesada}

\affil[1]{D\'epartement de g\'enie physique, \'Ecole polytechnique de Montr\'eal, Montr\'eal, QC, H3T 1J4, Canada}

\affil[*]{vincent-2.girouard@polymtl.ca}
\affil[$\dagger$]{nicolas.quesada@polymtl.ca}

\begin{abstract}
Universal multiport interferometers (UMIs) have emerged as a key tool for performing arbitrary linear transformations on optical modes, enabling precise control over the state of light in essential applications of classical and quantum information processing such as neural networks and boson sampling. 
While UMI architectures based on Mach-Zehnder interferometer networks are well established, alternative approaches that involve interleaving fixed multichannel mixing layers and phase masks have recently gained interest due to their high robustness to losses and fabrication errors. However, these approaches currently lack optimal analytical methods to compute design parameters with low optical depth. In this work, we introduce a constructive decomposition of unitary matrices using a sequence of $2N+5$ phase masks interleaved with $2N+4$ discrete Fourier transform matrices. This decomposition can be leveraged to design universal interferometers based on phase masks and multimode interference couplers, implementing a discrete Fourier transform, offering an analytical alternative to conventional numerical optimization-based designs and reducing by a factor of 3 the previous best known analytical methods.
\end{abstract}

\setboolean{displaycopyright}{false} 

\begin{document}

\maketitle

\section{Introduction}
Linear optics is a powerful framework for classical and quantum information processing, owing to the high transmission speed, low power consumption, and low decoherence of photons when used as information carriers. In addition, integrated photonics provides a mature and readily available platform for realizing these technologies in a compact, stable, and scalable manner \cite{wang2020integrated}.
A central component of such systems is the universal multiport interferometer (UMI), a reconfigurable optical device capable of performing arbitrary linear transformations on multiple optical channels \cite{harris2018linear, bogaerts2020programmable}.
UMIs play a crucial role in many key applications of quantum information processing such as boson sampling \cite{aaronson2011computational, spring2013boson, crespi2013integrated} and its variants~\cite{hamilton2017gaussian,deshpande2022quantum,grier2022complexity}, linear optical quantum computing \cite{knill2001scheme, kok2007linear, politi2008silica, baldazzi2025universal}, quantum simulations \cite{harris2017quantum, sparrow2018simulating, somhorst2023quantum}, quantum neural networks \cite{steinbrecher2019quantum,killoran2019continuous}, and continuous-variable quantum information processing~\cite{lloyd1999quantum,braunstein2005quantum,kalajdzievski2021exact,houde2024matrix,zhou2024bosehedral}. They are also crucial in classical applications, including fiber optic communication \cite{bozinovic2013terabit, choutagunta2019adapting}, sensing \cite{zelaya2025photonic}, signal processing \cite{zhuang2015programmable, perez2017multipurpose}, optical neural networks \cite{shen2017deep}, and imaging \cite{butaite2022build, kupianskyi2024all}. The development of compact, loss- and error-tolerant UMI designs is therefore critical for advancing those technologies.
 
\begin{figure}[!ht]
    \centering
    \includegraphics[width=\linewidth]{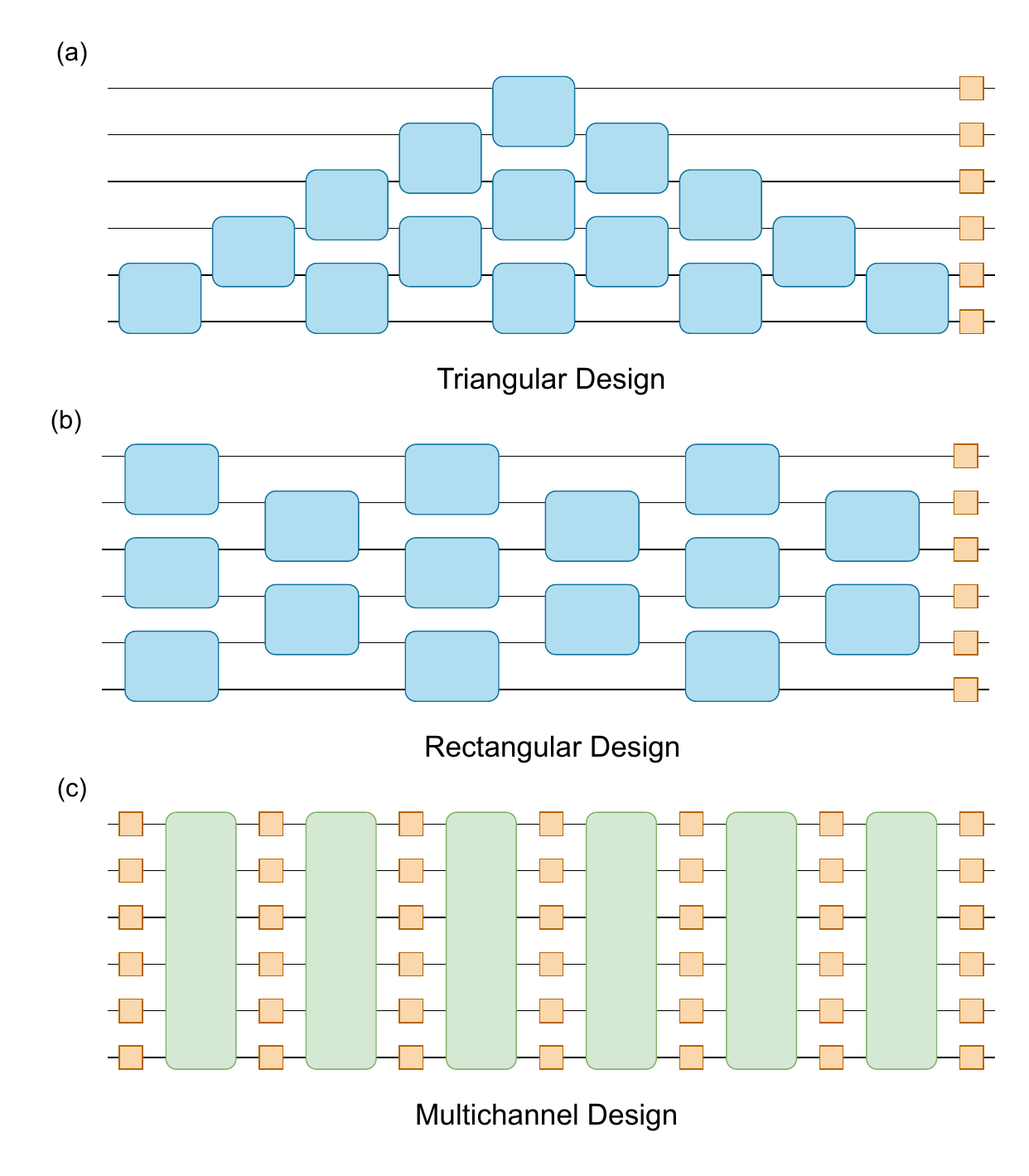}
    \caption{Different configurations of universal multiport interferometers. (a)~~Triangular mesh which implements the decomposition of Reck \textit{et al.}~\cite{reck1994experimental} in an integrated photonics setting~\cite{carolan2015universal} made of MZIs (blue rounded rectangles) and phase shifters (orange squares). (b)~~Rectangular mesh of Clements \textit{et al.}~\cite{clements2016optimal}. (c)~~Sequence of $L=N+1$ diagonal phase masks interleaved with a fixed multichannel mixing layer (green rounded rectangles).}
    \label{fig:configuration_review}
\end{figure}

The decomposition of unitary matrices into factors implementable by optical components plays a central role in the design of UMIs. 
It is well established that any unitary transformation can be implemented using specific arrangements of two-mode components \cite{Golub2013-js}.
In a seminal work, Reck \textit{et al.} \cite{reck1994experimental} showed that arbitrary linear transformations can be realized using a triangular network of Mach-Zehnder interferometers (MZI) and single-mode phase shifters, as shown in Fig.~\ref{fig:configuration_review}(a). This architecture enabled the realization of high-fidelity integrated optical processors for up to $N=6$ modes \cite{carolan2015universal}. This scheme was later improved by Clements \textit{et al.}  \cite{clements2016optimal}, who used the same unit cell in a rectangular mesh to achieve a more compact and loss-tolerant design, as depicted in Fig.~\ref{fig:configuration_review}(b). The Clements \textit{et al.} design has since been demonstrated for systems with up to $N=20$ modes \cite{taballione20198, taballione2021universal, taballione202320}. Other similar configurations have also been proposed to further enhance error and loss tolerance \cite{shokraneh2020diamond, mojaver2023addressing, fldzhyan2020optimal}, computational cost \cite{de2018simple}, or optical depth \cite{bell2021further}. However, scaling such architectures to a large number of modes remains a challenge due to the poor robustness of MZI networks to fabrication imperfections. Even small deviations in beam splitter reflectivities can significantly compromise universality of large UMIs \cite{burgwal2017using}, imposing strict requirements on component precision and increasing fabrication complexity. To mitigate these effects, various strategies are typically employed \cite{mower2015high, burgwal2017using, pai2019matrix, miller2013self_align, miller2013self_conf, miller2015perfect, miller2017setting, hamerly2022stability, hamerly2022accurate, bandyopadhyay2021hardware}. Some approaches introduce redundant layers or mesh non-localities \cite{mower2015high, burgwal2017using, pai2019matrix} and rely on global optimization to set MZI parameters. However, such methods increase the optical depth of the device, require pre-characterization of component errors, and are computationally expensive. Other strategies use self-configuring networks \cite{miller2013self_align, miller2013self_conf, miller2015perfect, miller2017setting, hamerly2022stability, hamerly2022accurate}, requiring additional sources and power detectors inside or outside the circuit, which further increases fabrication complexity.

To improve tolerance to losses and component imperfections, several approaches have explored alternative designs based on multichannel components  \cite{kumar2021unitary, arends2024decomposing, yasir2025compactifying,markowitz2025photonic}. Among them, architectures based on interleaving a fixed multichannel mixing layer with phase masks \cite{tang2017integrated, tang2021ten, tanomura2022scalable, saygin2020robust, lopez2021arbitrary, markowitz2023universal, pereira2025minimum} [see Fig.~\ref{fig:configuration_review}(c)] are of particular relevance because of their high robustness against perturbations in the mixing layer \cite{saygin2020robust, markowitz2023auto, tanomura2022scalable}. 
Parameter counting gives a lower bound of at least $N+1$ phase masks needed to represent an arbitrary unitary. This is consistent with numerical evidence suggesting that interleaving $N+1$ \cite{saygin2020robust, markowitz2023universal} or $N+2$ \cite{pereira2025minimum} layers of phase masks with almost any dense unitary matrix could result in a universal architecture \cite{zelaya2024goldilocks}. Notable examples of mixing layers include the discrete Fourier transform (DFT), which can be implemented using a multimode interference coupler \cite{lopez2021arbitrary, bachmann1994general}, and the discrete fractional Fourier transform (DFrFT), which can be realized using multiport waveguide arrays \cite{weimann2016implementation}. However, unlike designs based on MZI networks, those built from multichannel mixing layers lack optimal analytical algorithms for computing the phase mask parameters and thus almost always rely on numerical optimization procedures \cite{markowitz2023universal, saygin2020robust, pereira2025minimum, zelaya2024goldilocks}.
As a result, their universality 
cannot be guaranteed, and the high computational cost associated with computing parameters prevents their scalability to larger systems, highlighting the need for optimal analytic designs.

Significant progress has already been made toward an analytical decomposition of linear transformations into a sequence of phase masks and DFT operations. Tools from group theory have been used to provide an existence proof that all unitary operations can be constructed from alternating DFTs and phase masks \cite{borevich1981subgroups}, although the exact number of masks required in the sequence is not specified. Later works in matrix analysis have shown that any square matrix can be decomposed into a product of diagonal and circulant matrices \cite{schmid2000decomposing} and provided a constructive proof requiring at most $2N-1$ factors for an $N \times N$ matrix \cite{huhtanen2015factoring}. This result is particularly relevant since circulant matrices can be diagonalized by the DFT matrix \cite{kra2012circulant}. However, the algorithm does not guarantee that the diagonal matrices are unitary, which prevents them from being implemented as passive phase masks. Using the Sinkhorn normal form for
unitary matrices, Idel and Wolf provided a non-constructive proof that any $N\times N$ unitary matrix can be decomposed into $2N-1$ diagonal unitaries interleaved with partial DFTs \cite{idel2015sinkhorn}. However, their work does not provide a procedure for explicitly computing the phase mask parameters. 

The most promising approach toward the goal of decomposing unitary linear transformations into a sequence of phase masks and DFT operations was developed by López Pastor \textit{et al.} \cite{lopez2021arbitrary}, who introduced a constructive proof showing that any $N\times N$ unitary matrix can be decomposed into a product of $6N+1$ phase masks interleaved with DFT matrices. Their method first applies the algorithm of Clements \textit{et al.}~\cite{clements2016optimal} to express the unitary as a rectangular network of beam splitters and phase shifters, and then interprets each layer as a multimode interaction, which is subsequently decomposed into a product of circulant and diagonal matrices. However, this method generates a sequence with a significant phase mask overhead, as numerical evidence and simple parameter counting suggest that as few as $N+1$ layers could suffice \cite{saygin2020robust, zelaya2024goldilocks}.

In this work, we introduce a new analytical decomposition of unitary matrices in terms of DFTs and phase masks, which yields a universal multiport interferometer with only a third of the optical depth compared to previous analytical designs \cite{lopez2021arbitrary}. We also provide an open-source Python implementation of the algorithm.
Our approach builds on the framework of López Pastor \textit{et al.} \cite{lopez2021arbitrary}, but begins with the interferometer of Bell and Walmsley \cite{bell2021further}, which uses a symmetric MZI (sMZI) as unit cell instead of the asymmetric MZI used in the Clements \textit{et al.} design \cite{clements2016optimal}. This building block is not only more compact but also more symmetric, which results in significant simplifications in the sequence of phase masks. We also introduce a circuit identity to increase the symmetry of the layers, allowing additional compression of the overall sequence. This reduction in the number of layers could considerably lower optical losses, fabrication costs, and size of UMIs based on this architecture, thereby facilitating their scalability to a larger number of modes. 

\section{Decomposition Method}
A lossless linear transformation acting on $N$ optical modes can be represented by a $N \times N$ unitary matrix $U$. Such a matrix transforms the annihilation operators (or the classical amplitudes) of the input modes according to
\begin{align}
    a_j \rightarrow b_j = \sum_{k=0}^{N-1}U_{jk}a_k.
\end{align}
Arbitrary unitary matrices of size $N \times N$ are characterized by $N^2$ real degrees of freedom \cite{itzykson1966unitary}. Hence, any architecture aiming to implement arbitrary transformations needs at least the same amount of controllable parameters. This constraint puts a lower bound on the number of phase masks required in the design, as at least $N+1$ masks are necessary to ensure the complete parametrization of the unitary.
Without loss of generality, in what follows we assume that $N$ is even.

We want to find an analytical decomposition of $U$ such that
\begin{align}
    U = D^{(0)}FD^{(1)}FD^{(2)} \dots FD^{(L)},
    \label{eq:what_we_want}
\end{align}
where $L$ should be as small as possible, $D^{(k)}$ are diagonal phase masks of the form
\begin{align}
    D^{(k)} = \text{diag}\left\{e^{i\phi_0^{(k)}}, e^{i\phi_1^{(k)}}, \dots, e^{i\phi_{N-1}^{(k)}}\right\},
\end{align}
and $F$ is the DFT matrix~\cite{aristidou2007logarithm}, whose elements are given by
\begin{align}
    F_{j k} = \frac{1}{\sqrt{N}}e^{-i2\pi jk/N} \text{ for } i,j\in\{0,1,\ldots,N-1\}.
\end{align}
An important property of the DFT matrix, which will prove to be particularly useful, is that it diagonalizes circulant matrices \cite{kra2012circulant}. Circulant matrices have each row given by a cyclic right shift of the previous one. If $C$ is circulant, then there exists a diagonal matrix $D$ such that $C = F^\dagger D F$. The problem of finding a decomposition of $U$ as in~\eqref{eq:what_we_want} is therefore equivalent to finding a decomposition of $U$ in terms of diagonal and circulant unitary matrices.

\begin{figure*}[htbp]
    \centering
    \includegraphics[width=\linewidth]{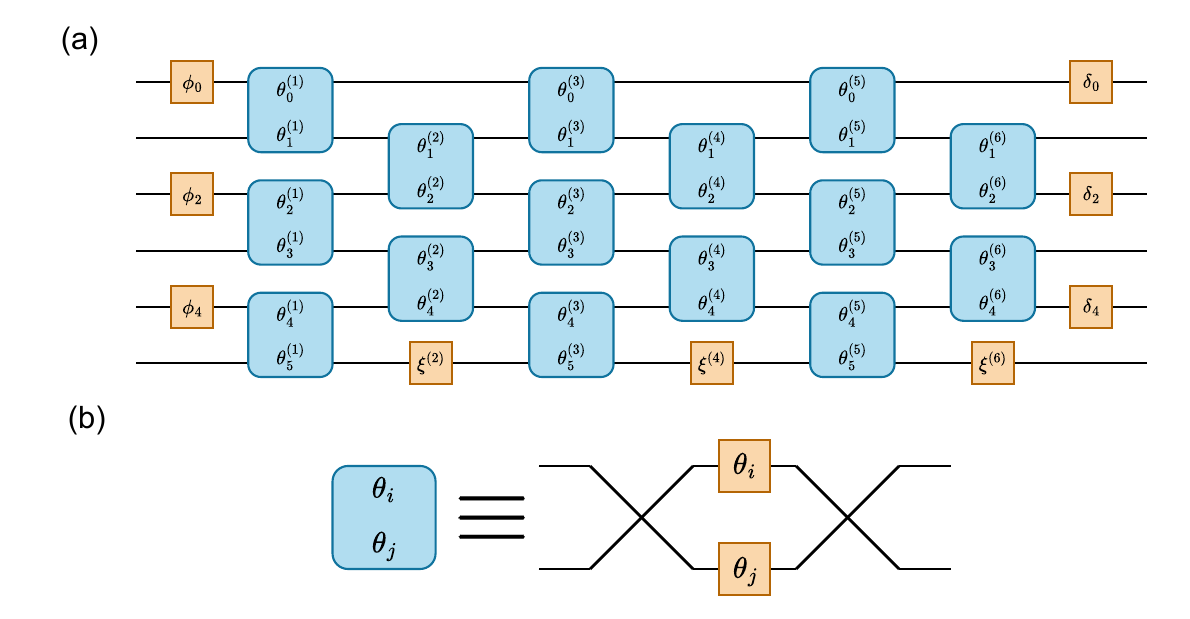}
    \caption{Diagram of the Bell and Walmsley universal multiport interferometer \cite{bell2021further}. (a) Layout of the interferometer for $N=6$ modes, where blue rounded rectangles correspond to symmetric MZIs and orange squares correspond to single-mode phase shifters. A distinctive feature of this design is the presence of additional edge phase shifters in even MZI layers. (b) Unit cell of the interferometer. The unit cell consists of a sMZI made from two 50:50 beam splitters and two internal phase shifters.
    \label{fig:Bell_design}}
\end{figure*}

Bell and Walmsley~\cite{bell2021further} demonstrated that any $N \times N$ unitary matrix can be decomposed into a rectangular mesh of $N(N-1)/2$ symmetric MZI (sMZI) with additional phase shifters located along the edges of the mesh. 
An example of this architecture is shown in Fig.~\ref{fig:Bell_design}(a) for $N=6$ modes. Their design is characterized by the use of a symmetric non-universal unit cell, which is more compact than those employed in~\cite{reck1994experimental, clements2016optimal}, resulting in an interferometer with reduced optical depth. The sMZI used in this design is made from two balanced beam splitters and two internal phase shifters, as depicted in Fig.~\ref{fig:Bell_design}(b). It acts on two consecutive modes $m$ and $n$ according to the transfer matrix $T_{m,n}(\theta_m, \theta_n)$ given by
\begin{align}
    T_{m, n}(\theta_m, \theta_n) &= X \Theta X = e^{i\Sigma}\begin{pmatrix}
        \cos{\delta} && i\sin{\delta}
        \\ i\sin{\delta} && \cos{\delta}
    \end{pmatrix},
    \label{eq:transfert_matrix}
\end{align}
where $\Theta = \text{diag}\left\{e^{i\theta_m}, e^{i\theta_n}\right\}$, $\Sigma = \left(\theta_m + \theta_n\right)/2$, $\delta = \left(\theta_m - \theta_n\right)/2$, and $X$ corresponds to the 50:50 beam splitter matrix given by 
\begin{align}
    X = \frac{1}{\sqrt{2}}\begin{pmatrix}
        1 && 1\\
        1 && -1
    \end{pmatrix}.
\end{align}
When the unit cell operates within a multimode system ($N > 2$), the resulting transfer matrix is the $2\times2$ matrix defined in ~\eqref{eq:transfert_matrix} embedded into the $N \times N$ identity matrix at modes $m$ and $n$. 

\begin{figure}[htbp]
    \centering
    \includegraphics[width=\linewidth]{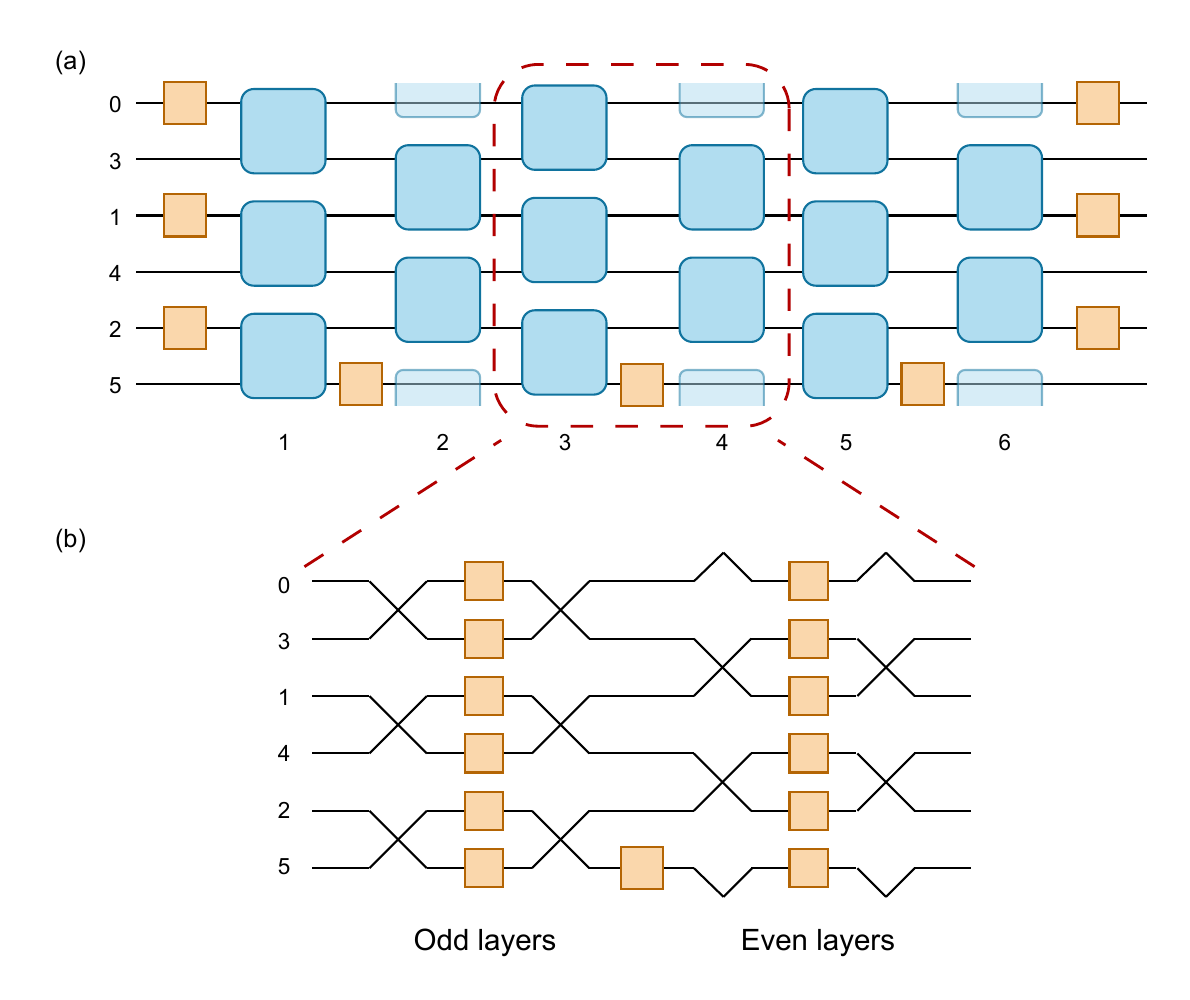}
    \caption{Modified multiport interferometer used to derive the mask sequence. (a) Interactions are added between the first and last channel to simulate a periodic system. Channels are also relabelled using the permutation matrix of ~\eqref{eq:k_permutation_matrix}. (b) Structure of one bi-layer. In the odd layer, sMZIs generate interactions between channels $j$ ($j \geq N/2$) and $j-N/2$. In even layers, sMZIs generate interactions between channels $j$ ($j \geq N/2$) and $j-N/2+1$.
    Note that the sMZIs connecting modes 0 and $N-1$ are set to the bar (identity) configuration.
    }
    \label{fig:bi-layer}
\end{figure}

We will interpret each sMZI layer in the interferometer from Fig.~\ref{fig:Bell_design} as a multimode interaction. To ensure that each layer acts uniformly on the entire mode space, we will impose periodic boundary conditions on the mode indices so that interactions between channels $j$ and $j+1$ (mod $N$) will occur in even-numbered layers, as shown in Fig.~\ref{fig:bi-layer}(a). The fictitious interaction between modes 0 and $N-1$ happening at the boundary is set to the bar state (identity matrix), but is included for completeness. We will also follow the procedure of \cite{lopez2021arbitrary} and relabel the channels of the interferometer such that $\left\{0, 1, 2, \dots, N-1\right\} \rightarrow \left\{0, \frac{N}{2}, 1, \frac{N}{2} + 1, \dots, N-1\right\}$. This transformation can be achieved using the permutation matrix $K$, defined by 
\begin{align}
    K_{j k} = \begin{cases}
        1 \qquad k=2j,\ j \leq \frac{N}{2} - 1\\
        1 \qquad k = 2j+1-N,\ j > \frac{N}{2}-1\\
        0 \qquad \text{otherwise}
    \end{cases}.
    \label{eq:k_permutation_matrix}
\end{align}
The effect of this permutation on the channel labels can be seen in Fig.~\ref{fig:bi-layer}(a) for $N=6$ channels.

With this relabeling, odd sMZI layers exhibit a simple structure, where each channel $j$ of the second half of the system (i.e., $j \geq N/2$) interacts with channel $j-N/2$, as highlighted in Fig.~\ref{fig:bi-layer}(b). This structure allows the transformation performed by an odd layer $k$ to be described by 
\begin{align}
    T_\text{odd}^{(k)} = X\Theta^{(k)}X.
    \label{eq:odd_layers}
\end{align}
In ~\eqref{eq:odd_layers}, $X$ represents a layer of $N/2$ beam splitters acting between pairs of channels $j$ ($j \geq N/2$) and $j - N/2$. The corresponding transformation is given by the block matrix
\begin{align}
    X = \frac{1}{\sqrt{2}}\begin{pmatrix}
        I && I\\
        I && -I
    \end{pmatrix},
    \label{eq:x_matrix}
\end{align}
where $I$ is the $\frac{N}{2} \times \frac{N}{2}$ identity matrix. $\Theta^{(k)}$ is a diagonal matrix that represents a layer of phase shifters applied to all modes and is given by
\begin{align}
    \Theta^{(k)} = \text{diag}\Bigl\{ &e^{i\theta_{0}^{(k)}}, e^{i\theta_{2}^{(k)}}, \dots, e^{i\theta_{N-2}^{(k)}}, \nonumber \\
    &e^{i\theta_{1}^{(k)}},
    e^{i\theta_{3}^{(k)}}, \dots, e^{i\theta_{N-1}^{(k)}}\Bigr\} .
\end{align}

In even layers, all unit cells are shifted by one, resulting in interactions between each channel $j$ ($j\geq N/2$) and $j-N/2 + 1$ [see Fig.~\ref{fig:bi-layer}(b)]. The transformation can be described by the same sequence as \eqref{eq:odd_layers}, provided that a channel permutation is applied beforehand. As detailed in \cite{lopez2021arbitrary}, a cyclic shift must be applied to the first half of the modes to map each channel $j-N/2 + 1$ (mod $N/2$) ($j \geq N/2$) to $j-N/2$, thereby restoring the structure observed in odd layers thanks to the invariance of sMZIs under channel exchange. This transformation can be implemented using the permutation matrix $P$, defined as
\begin{align}
    P_{jk} = \begin{cases}
        1 \qquad k=j+1 \left(\text{mod}\ \frac{N}{2}\right), \quad j\leq \frac{N}{2} - 1\\
        1 \qquad k=j, \quad j>\frac{N}{2}- 1\\
        0 \qquad \text{otherwise}
    \end{cases}.
    \label{eq:p_matrix}
\end{align}
This permutation cyclically shifts by one the first half of the modes while leaving the second half unchanged.
We can thus express the transformation performed in even layers by
\begin{align}
    T_\text{even}^{(k)} = P^TX\Omega^{(k)}XP,
    \label{eq:even_layers}
\end{align}
where 
\begin{align}
    \Omega^{(k)} = \text{diag}\Bigl\{&e^{i\theta_{1}^{(k)}}, e^{i\theta_{3}^{(k)}}, \dots, e^{i\theta_{N-1}^{(k)}}, \nonumber \\
    &e^{i\theta_{2}^{(k)}}, e^{i\theta_{4}^{(k)}}, \dots, e^{i\theta_{0}^{(k)}}\Bigr\}.
\end{align}

The unitary matrix $U$ can therefore be expressed in an intermediate form as 
\begin{align}
    U = DK^T\left(\prod_{k=1}^{N/2}T_\text{even}^{(2k)}\Xi^{(2k)}T_\text{odd}^{(2k-1)} \right)KD',
    \label{eq:intermediate_form}
\end{align}
where $T_\text{odd}$ and $T_\text{even}$ are given by ~\eqref{eq:odd_layers} and \eqref{eq:even_layers}. The matrices $D'$ and $D$ are the diagonal phase masks applied at the input and output of the circuit [see Fig.~\ref{fig:Bell_design}], while $\Xi^{(k)} = \text{diag}\left\{1, 1, \dots, e^{i\xi^{(k)}}\right\}$ corresponds to the edge phase shifters present in the Bell and Walmsley architecture \cite{bell2021further}. In the decomposition \eqref{eq:intermediate_form}, each recurring sMZI bi-layer takes the form
\begin{align}
    T_\text{bi-layer} = T_\text{even}\Xi T_\text{odd} = P^TX\Omega XP\Xi X\Theta X,
    \label{eq:bi_layer}
\end{align}
where the layer indices have been omitted for clarity.

López Pastor \textit{et al.} \cite{lopez2021arbitrary} showed that the matrix $P$ in ~\eqref{eq:p_matrix} can be decomposed as
\begin{align}
    P = X\Lambda X,
    \label{eq:p_decomposition}
\end{align}
where $\Lambda = F^\dagger H F$ is circulant, and $H$ is a diagonal matrix with entries
\begin{align}
    H_{jj} = \frac{1}{2}\left[1 - (-1)^j\right] + \frac{1}{2}\left[1 + (-1)^j\right]e^{i2\pi j / N}.
    \label{eq:h_matrix}
\end{align}
By substituting \eqref{eq:p_decomposition} into \eqref{eq:bi_layer}, and using the fact that $XX = I$, we obtain
\begin{align}
    T_\text{bi-layer} = X\left[\Lambda^T\Omega \Lambda X\Xi X \Theta\right]X,
    \label{eq:bi_layer_2}
\end{align}
where the square brackets indicate that the two $X$ matrices at the edges will cancel out for all but the first and last bi-layers. 
\begin{figure}[htbp]
    \centering
    \includegraphics[width=\linewidth]{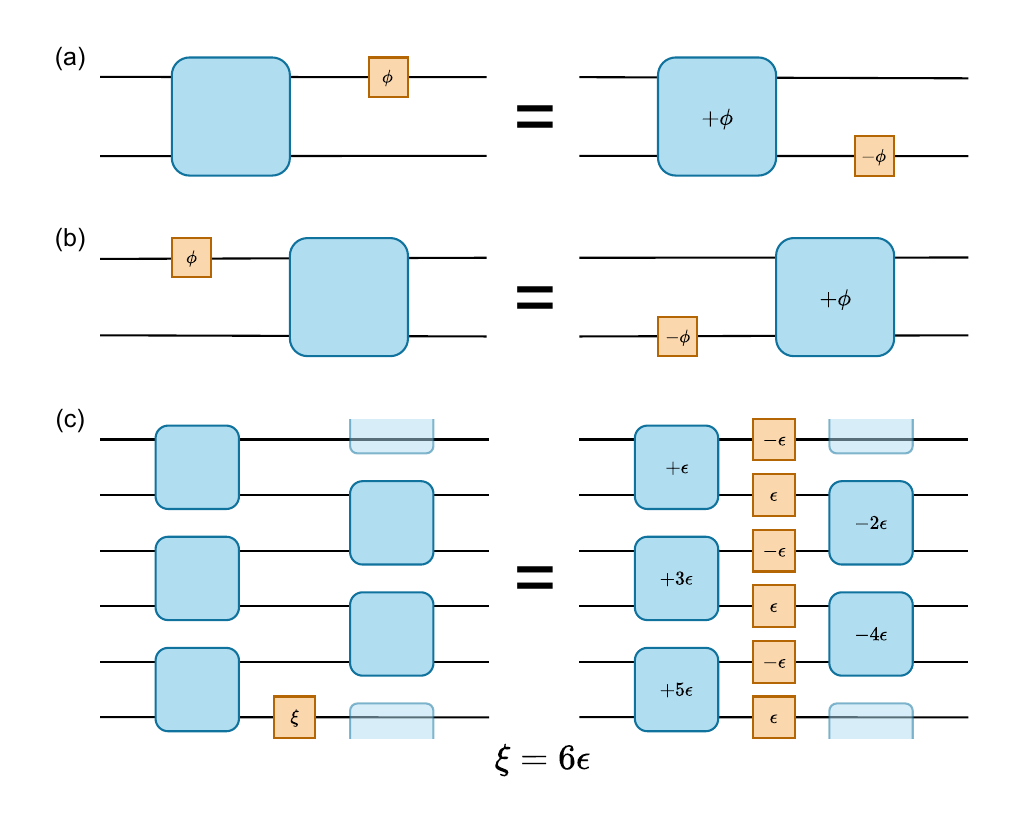}
    \caption{Relocation of edge phase shifters in the interferometer. (a - b)~~In a sMZI network, phase shifters can be moved to adjacent modes by changing their sign and adjusting the global phase of the neighboring sMZI. (c)~~The edge phase shifter can be divided into $N$ equal partial phases that can be distributed to all channels of the interferometer.}
    \label{fig:phase_relocation}
\end{figure}

The expression \eqref{eq:bi_layer_2} can be further simplified by using circuit identities to relocate the edge phases within the interferometer. In the Bell and Walmsley design \cite{bell2021further}, edge phases can be moved to different modes by adjusting the global phase of the adjacent sMZI, as illustrated in Figs.~\ref{fig:phase_relocation}(a) and \ref{fig:phase_relocation}(b). This principle can be leveraged to distribute evenly the edge phases across all the modes in the interferometer, as explained in Fig.~\ref{fig:phase_relocation}(c). Thus, given an initial edge phase shift of angle $\xi$, the matrix $\Xi = \text{diag}\left\{1, 1, \dots, e^{i\xi}\right\}$ becomes
\begin{align}
    \Xi = \begin{pmatrix}
        Ie^{-i\epsilon} && 0\\
        0 && Ie^{i\epsilon}
    \end{pmatrix},
\end{align}
where $\epsilon = \xi/N$. Using this new symmetric form for $\Xi$, it can be seen that the product $X\Xi X$ in ~\eqref{eq:bi_layer_2} becomes circulant, i.e.,
\begin{align}
    X\Xi X = \begin{pmatrix}
        I\cos{\epsilon} && -Ii\sin{\epsilon}\\
        -Ii\sin{\epsilon} && I\cos{\epsilon}
    \end{pmatrix}.
    \label{eq:x_xi_x_circulant}
\end{align}
The matrix in ~\eqref{eq:x_xi_x_circulant} can therefore be diagonalized by DFTs into the $C$ matrix, whose entries are given by
\begin{align}
    C_{jj} = e^{i(-1)^{j+1}\epsilon}.
\end{align}

After these simplifications, all circulant matrices in ~\eqref{eq:bi_layer_2} can be diagonalized, which results in
\begin{align}
    T_\text{bi-layer} = X\left[F H F^\dagger\Omega F^\dagger HCF\Theta\right]X.
    \label{eq:bi_layer_3}
\end{align}
An important property of the DFT matrix is that $F^4 = I$, which means that $F^\dagger = F^3 = \Pi F = F\Pi$ where $\Pi$ is a permutation matrix with elements given by
\begin{align}
    \Pi_{j k} = \begin{cases}
        1\qquad j=k=0\\
        1\qquad j=N-k\\
        0\qquad \text{otherwise}
    \end{cases}.
    \label{eq:pi_matrix}
\end{align}
It follows that all daggers can be removed from ~\eqref{eq:bi_layer_3} to get 
\begin{align}
    T_\text{bi-layer} = X\left[F HFp\left(\Omega\right)F HCF\Theta\right]X,
    \label{eq:bi_layer_final}
\end{align}
where $p:U \rightarrow\Pi U\Pi$. Using ~\eqref{eq:bi_layer_final} in expression \eqref{eq:intermediate_form} for $U$, we get

\begin{align}
    U = &DK^TX \nonumber\\
    \times&\left(\prod_{k=1}^{N/2}F HFp\left(\Omega^{(2k)}\right)F HC^{(2k)}F\Theta^{(2k-1)}\right) \nonumber\\
    \times &XKD'.
    \label{eq:final_expression_with_x}
\end{align}

The final step is to decompose the two remaining $X$ matrices in ~\eqref{eq:final_expression_with_x}. López Pastor \textit{et al.}~\cite{lopez2021arbitrary} showed that $X$ can be written as 
\begin{align}
    X = GYG,
    \label{eq:x_decomp}
\end{align}
where
\begin{align}
    G = \begin{pmatrix}
        I && 0\\
        0 && iI
    \end{pmatrix}
    \label{eq:g_matrix}
\end{align}
is diagonal, and
\begin{align}
    Y = F^\dagger E F,
    \label{eq:y_matrix}
\end{align}
is circulant, with the matrix $E$ given by
\begin{align}
    E_{jj} = \frac{1}{\sqrt{2}}\left[1 - i(-1)^j\right].
\end{align}
~\eqref{eq:x_decomp} and \eqref{eq:y_matrix} can thus be used in \eqref{eq:final_expression_with_x} to obtain the final expression for $U$:
\begin{align}
    U = &K^T\Pi \Bigl[p\left(k(D)G\right)FEFG \nonumber\\
    \times &\left(\prod_{k=1}^{N/2}F HFp\left(\Omega^{(2k)}\right)F HC^{(2k)}F\Theta^{(2k-1)}\right) \nonumber\\
    \times &GFEFp\left(Gk(D')\right)\Bigr]\Pi K,
    \label{eq:final_decomposition}
\end{align}
where $k: U \rightarrow KUK^T$.
Decomposing $U_p = K^T\Pi U\Pi K$ yields an expression for $U$ entirely in terms of DFT and diagonal matrices. The decomposition \eqref{eq:final_decomposition} consists of a sequence of $2N+5$ phase masks and $2N+4$ DFT matrices for an $N \times N$ unitary matrix. Only the matrices with upper indices depend on $U$, while the others are fixed. Although the proposed derivation only works in the case of unitary matrices of even dimension, odd unitaries can always be embedded into a larger even-dimensional identity matrix at the cost of introducing two additional phase masks in the final sequence.
\section{Discussion}
The universal design proposed in this work represents a significant improvement in mask overhead compared to the previous analytical approach, which required $6N+1$ masks \cite{lopez2021arbitrary}. The final sequence obtained in ~\eqref{eq:final_decomposition} consists of a product of $2N+5$ phase masks interleaved with $2N+4$ DFT layers, representing a 66\% reduction in optical depth for large $N$. Our main contribution over this previous design lies in the use of a multiport interferometer built from symmetric MZIs. Thanks to their more compact layout, all phase shifters within a given sMZI layer are aligned in parallel, enabling highly parametrized phase masks and helping reduce the overall circuit depth. By applying circuit identities, localized edge phases were uniformly distributed across all modes, leading to more convenient and symmetric matrix representations. Interestingly, by relocating a single component in the unit cell — specifically the external phase shifter, which accounts for half of the length taken by phase shifters in the Clements \textit{et al} \cite{clements2016optimal} interferometer — we are able to divide the overall depth by a factor of 3 when using DFT layers and phase masks.

This decomposition has the potential to enable the design of UMIs that are more robust, compact, cheap, and that exhibit lower propagation losses. These improvements could facilitate the scaling of photonic processors to a larger number of modes, allowing more complex circuits to be integrated on a single photonic chip. Such advancements would benefit the development of both classical and quantum photonic technologies. 
Although a full analysis of imperfections in our decomposition lies outside the scope of this work, we would like to highlight that it is highly symmetric, and thus if each element, DFT and phase mask, has identical transmission properties then losses will be balanced for the whole interferometer, a highly desirable feature.
The analytic method that was developed also provides a fast and exact way to compute the mask parameters without relying on numerical optimization, thereby enabling faster and more energy-efficient programming of UMIs, as the time complexity of this approach matches that of the Clements \textit{et al.} or Bell and Walmsley schemes \cite{clements2016optimal, bell2021further}. 
Additionally, this design is also resilient to noise and fabrication errors since perturbations in the mixing layer do not compromise universality \cite{markowitz2023auto}, eliminating the need for redundant layers. Moreover, we expect that an analytical framework could help develop faster and more accurate error correction strategies in the presence of fabrication imperfections.

We note that the derivation introduced in this work gives rise to a continuous family of new analytical decompositions of a unitary matrix as sequences of mixing layers and phase masks. Specifically, many complex Hadamard matrices that are equivalent to the $N \times N$ DFT matrix can serve as valid mixing layers. Given a permutation matrix $P$ and two diagonal unitaries $\Lambda_1$ and $\Lambda_2$, we can define a generalized mixing layer as $\tilde{F} = \Lambda_1PFP^T\Lambda_2$. By appropriately adjusting the phase masks such that $\tilde{D}^{(k)} = \Lambda_2^{*}PD^{(k)}P^T\Lambda_1^{*}$, they will remain diagonal and the structure of the decomposition will be preserved. The method could therefore be used with other physical implementations of mixing layers that are not represented by the DFT matrix.

While this decomposition is of particular relevance for photonics, with applications ranging from boson sampling and quantum simulations to classical signal processing, it could also be applied to any kind of linear wave, enabling programmable wave evolution in a broader context. More generally, our method introduces a simple and constructive matrix factorization relying only on diagonal unitary matrices and discrete Fourier transforms, or equivalently, on diagonal and circulant unitary matrices. Despite these promising advances, the number of phase masks required in this approach remains roughly twice the expected lower bound of $N+1$ that can be observed as a phase transition in many numerical experiments. This means that our design is still suboptimal in the number of layers. Achieving this lower bound remains an open challenge, but we expect that our work represents a significant step toward that goal.

{\bf{Disclosures --- }}
The authors declare no conflict of interest.

{ \bf Software Implementation ---}
A Python implementation of our algorithm, as well as the ones of Clements \emph{et al.} \cite{clements2016optimal}, Bell and Walmsley \cite{bell2021further} and L\'opez Pastor, Lundeen and Marquardt \cite{lopez2021arbitrary} is available in the Unitary-Decomp package~\cite{girouard2025}. The package can also perform gradient-based numerical decompositions of unitaries using JAX~\cite{jax2018github}.

{\bf Acknowledgments --- }
The authors thank the Ministère de l’Économie et de
l’Innovation du Québec and the Natural Sciences and
Engineering Research Council of Canada  (Quantum Consortium Program (Quantamole)) for their financial support and M. Trudeau for insightful discussions.

\bibliography{sample.bib}


\end{document}